\documentclass[12pt]{article}
\usepackage{epsfig}
\usepackage{graphicx}
\usepackage{cite}
\usepackage{amsfonts}
\usepackage{amsmath}
\usepackage{amssymb}
\usepackage{axodraw}
\usepackage{latexsym}
\def\ee{\end{equation}}
\def\ba{\begin{eqnarray}}
\def\ea{\end{eqnarray}}

\def\bq{\begin{quote}}
\def\eq{\end{quote}}

 at 10truept

\newcommand{\beq}{\begin{equation}}
\newcommand{\eeq}{\end{equation}}
\newcommand{\beqa}{\begin{eqnarray}}
\newcommand{\eeqa}{\end{eqnarray}}
\newcommand{\bea}{\begin{eqnarray}}
\newcommand{\eea}{\end{eqnarray}}
\newcommand{\p}{\partial}
\newcommand{\al}{\alpha}
 
 \newcommand{\ep}{\delta}

\newcommand{\bra}{\left<}
\newcommand{\ket}{\right>}

\newcommand{\overskrift}[1]{\vspace{6.0mm}\noindent\textbf{#1}\vspace{1.5mm}}

\def\lesssim{~\mbox{\raisebox{-.6ex}{$\stackrel{<}{\sim}$}}~}

\def\ltap{\ \raise.3ex\hbox{$<$\kern-.75em\lower1ex\hbox{$\sim$}}\ }
\def\gtap{\ \raise.3ex\hbox{$>$\kern-.75em\lower1ex\hbox{$\sim$}}\ }
\def\gl{\ \raise.5ex\hbox{$>$}\kern-.8em\lower.5ex\hbox{$<$}\ }
\def\roughly#1{\raise.3ex\hbox{$#1$\kern-.75em\lower1ex\hbox{$\sim$}}}

\newcommand{\picb}[1]{\;\parbox[c]{45pt}{\begin{picture}(45,30)(0,0)
\SetWidth{1.0}\SetScale{1.0} #1 \end{picture}}\;}

\begin{document}

\thispagestyle{empty}
\noindent
\begin{flushright}
CERN-PH-TH/2008-006
\end{flushright}

\vskip2cm
\begin{center}
{\Large{\bf On Resumming Inflationary Perturbations}}
\vskip0.2cm 
{\Large{\bf beyond One-loop}}
\vskip2cm {\large Antonio Riotto$^{1,2}$\footnote{\tt 
antonio.riotto@cern.ch} 
and Martin S. Sloth$^3$\footnote{\tt sloth@phys.au.dk}}\\

\vspace{.5cm}

\vskip 0.1in

{\em $^1$ CERN, Theory Division, Gen\'eve 23, CH-1211 Switzerland\\}

\vskip 0.1in
{\em $^2$ INFN, Sezione di Padova, Via Marzolo,
8 - I-35131 Padua -- Italy\\}
\vskip 0.1in
{\em $^3$ Department of Physics and Astronomy, University of Aarhus}\\ 
{\em DK-8000 Aarhus C, Denmark}\\

\vskip 0.1in
\vskip 0.1in
\vskip .25in
{\bf Abstract}
\end{center}
It is well known that the correlation functions of a  scalar field  in a quasi-de Sitter space exhibit at the loop level  
cumulative infra-red effects proportional to the total number of e-foldings of inflation. Using the in-in formalism, we explore 
the behavior of these infra-red effects in the large $N$ limit of an $O(N)$ invariant scalar field theory with quartic 
self-interactions. By resumming all higher-order loop diagrams non-perturbatively, we show that the connected four-point correlation 
function, which is a signal of  non-Gaussianity, 
is non-perturbatively enhanced with respect to its tree-level value.


\vfill \setcounter{page}{0} \setcounter{footnote}{0}
\newpage

\setcounter{equation}{0} \setcounter{footnote}{0}

\section{Introduction}
\noindent
In the inflationary picture the primordial cosmological perturbations are created from quantum 
fluctuations which are
``redshifted'' out of the horizon during an early period of 
accelerated expansion \cite{lrreview}.
These super-horizon perturbations  remain ``frozen''  (unless isocurvature fluctuations are
present) until the horizon
grows during a later matter- or radiation-dominated era.
After falling back inside the horizon they are communicated to the primordial
plasma and hence are directly
observable as temperature anisotropies in the Cosmic Microwave Background (CMB) radiation. 

Given the spectacular accuracy by which these anisotropies 
are and will be mapped  by 
present and forthcoming satellite 
experiments, the issue of  non-Gaussianity (NG) in the cosmological
perturbations has becoming more and more relevant \cite{review}.
Such non-Gaussianities are sourced by  self-interactions, including the ones 
induced by gravity,  in the early universe,
and become manifest
at the level of second- or higher-order perturbation theory.
Furthermore, and more important for the scope of   this  paper, 
self-interactions of any scalar field during the inflationary stage
give rise to  corrections in all correlators,
including the observationally interesting cases of the
two- and the three-point correlation functions.
Such  corrections are associated with so-called ``loops,''
in which virtual particles with arbitrary momentum are emitted and
re-absorbed by the fields which participate in the correlation function.

While loop corrections may lead to significant backreaction effects
\cite{Mukhanov:1996ak,Abramo:1997hu,boy1,boy2,olandesi}
which scale like (powers of)  
the number of e-folds between horizon exit of
the mode $k$ under consideration and the end of inflation, in this paper
we will concern ourselves with another aspect of these loop corrections. 
Indeed, it has been known for a long time that there exists a breakdown
in the perturbative
expansion due to infra-red (IR) ``divergences'' \cite{sasaki1,sasaki2}. 
These  classical one-loop corrections scale like  $\ln(k L)$ where 
$L^{-1}$ is the IR comoving momentum cut-off.
These IR divergences
have attracted lot of attention in the  literature recently, 
where in particular the loop 
corrections to cosmological correlation functions has been 
studied \cite{Sloth:2006az,Sloth:2006nu,Weinberg:2005vy,wein2,seery1,seery2,Urakawa:2008rb}. 
At the generic $n$-th order of perturbation theory, the power spectrum
of the perturbations 
is expected to get corrections growing like $\ln^n(k L)$, signalling the 
breakdown
of perturbation theory unless the corresponding coupling constant is tiny 
enough.

 One option is 
to set $L$ to be some comoving length scale which left the
horizon many e-folds before the observable universe (super-large box). In particular,
the smallest possible value of $L^{-1}$ is  
$a_i H$, where $a_i$ is the value of the scale factor at the beginning
of inflation. Since the wavelength
$k^{-1}$ goes out of the horizon when the scale factor equals
$a_k=k/H$, we see that 
$\ln(k L)=\ln(a_k/a_i)$, which is proportional to the total
number of e-folds from the beginning of inflation to the time when the
mode $k$ exits the horizon. This is the large cumulative 
IR correction as it gets larger the longer inflation
lasts. Indeed it was estimated that in models of chaotic inflation, where the total number of e-folding is large, the one-loop correction to the power spectrum can be significant \cite{Sloth:2006az,Sloth:2006nu,seery1,seery2,Urakawa:2008rb}. This may have important indirect implications for our understanding of the relation between observational quantities such as the tensor-to-scalor relation and the underlying fundamental model \cite{Sloth:2006az,Sloth:2006nu,Lyth:2007jh,seery1,seery2,bmprs,Urakawa:2008rb}.

 If we  insist in adopting a superlarge box to cut-off the IR divergences, it should be kept in mind
however that the location of  the much smaller box which includes our observable universe  
may be untypical and one should quantify how likely it is that the correlators
averaged over the superlarge box coincide with the correlators in
our observable universe \cite{lythbox,bmprs}. 
To deal with this problem, one can try to use the approach of stochatic
inflation \cite{starob} where the uncertainty in
the prediction coming from the large IR contributions in the superlarge box is traded for
the uncertainty inherent in having a probability distribution
for the background quantities \cite{bmprs}. 
However, before attacking the problem of the IR divergences from this point of view, we should first
ask ourselves whether it is possible to deal with them using the traditional 
quantum field theory techniques. 
We  may try to resum them, for example by using the
standard method of the Wilsonian Renormalization Group (RG). 
Unfortunately, this technique  is not very   efficient in practice.   
Indeed, the full set of exact RG equations
need to be solved. Approximations inevitably lead to disregarding a
set of diagrams which are not at all subleading in the IR. We refer the reader
to the Appendix of Ref.\cite{bmprs} for more details. 

We are though aware of one quantum field theory example in which the resummation
of the IR divergences may be performed, that is the case of $N$ scalar fields subject to an $O(N)$
symmetry and with a quartic interaction in which the free parameter $N$ is taken to be large.
This toy-model has been proved to be very useful  in dealing with  similar problems of IR resummation of perturbation
series in the case of finite temperarure field theory \cite{Dolan:1973qd,altherrfinite} and non-equilibrium field
theory \cite{altherr,rt}. We will then review how in the large $N$ limit the 
$O(N)$ theory embedded
in de Sitter can be solved and that the IR divergences can be resummed. Through the
in-in formalism \cite{sk},  one is able to give the expression for the 
exact two-point correlation function of the perturbations of the $N$ scalar fields and to compute
the exact connected four-point correlation function, dubbed trispectrum, 
which signals
the presence of NG. The resummed trispectrum will show a parametric dependence
on the quartic coupling which differs from the tree-level one. More important, the
correlation functions will not manifest IR divergences. Despite the fact that we will be dealing with a toy model, the results obtained in this paper may be useful to understand the behaviour
of more realistic systems in the IR.

The resummation of IR divergences was previously discussed by Hu and O' Connor in the 2PI (Two-Particle-Irreducible) formalism \cite{Hu:1986cv}, which we review in section 2 and 3, and by Starobinsky and Yokoyama using stochastic field theory methods \cite{Starobinsky:1994bd}, the results of which we also review in section 3. 

Thus, the paper is organized as follows. In Section 2 we review the details of the model
and the in-in formalism. In Section 3 we review how to solve the toy model in the large $N$ limit. In Section
4 we present the nonperturbative computation of the trispectrum. In Section
5 we compute the correction to the comoving curvature perturbation coupled to $N$ scalar 
fields, which is the toy model studied by Weinberg in Ref.\cite{wein2}. Finally, our conclusions
are given in Section 6.

\section{The $O(N)$ invariant model in the in-in formalism}

We consider a $O(N)$ invariant generalization of a scalar field theory with quadratic self-interaction of the type $\lambda\phi^4$. The generalized $O(N)$ invariant field theory consist of $N$ scalar fields $\phi_i$, $i=1,\dots,N$, with an action which is invariant under the $N$-dimensional real orthogonal group
\beq
\mathcal{L}= \frac{1}{2} \left(\p_{\mu}\phi_i\p^{\mu}\phi_i -m^2\phi_i\phi_i-\frac{\lambda}{4N}(\phi_i\phi_i)^2\right)~,
\eeq
where summation over repeated indices is assumed.
We will later embed this theory in an expanding universe. 
In the in-in formalism, the generating functional, $Z$, can be written
\bea
& &Z[J_+,J_-,\rho(t_{in})] = \\ \nonumber
& &\prod_i 
\int_{ctp}\mathcal{D}{\phi_i}^+\mathcal{D}{\phi_i}^-\exp\left[i\int_{t_{in}}^t dt' \int d^3x \sqrt{-g} (\mathcal{L}[{\phi_i}^+]-\mathcal{L}[{\phi_i}^-]+{J_i}_+{\phi_i}^+-{J_i}_-{\phi_i}^-)\right]~.
\eea
The closed time path (CTP) is defined by the condition $\phi_i^+(t)=\phi_i^-(t)$ and the boundary conditions in the asymptotic infinite past.
Now, by the usual Gaussian integration, one can write the free field generating functional in terms of the propagators
\beq
{{\bf{G}}_0}_{ij}(x,y) =\left(\begin{array}{cc} {G_0}_{ij}^{++}(x,y) & {G_0}_{ij}^{+-}(x,y) \\ {G_0}_{ij}^{-+}(x,y)&  {G_0}_{ij}^{--}(x,y)\end{array}\right) 
\eeq
such that for the free field generating functional, one then has
 \beq
 Z_0 \propto \exp\left[-i\frac{1}{2}\sum_{ij}\textbf{J}_i^T \textbf{G}_{ij}\textbf{J}_j\right] ~,
 \eeq 
where ${\bf J}_i^T=(J_i^+,J_i^-)$ and the Green functions are given by
\begin{align}
{G_0}^{-+}_{ij}(x,y) &=  i \langle \phi_i(x) \phi_j(y) \rangle^{(0)},
\label{gmp} \\ 
{G_0}^{+-}_{ij}(x,y) &=  i \langle \phi(y)_j \phi(x)_i \rangle^{(0)},
\label{gpm} \\ 
{G_0}^{++}_{ij}(x,y) &=  i \langle {\rm T} \phi(x)_i \phi(y)_j \rangle^{(0)} =
\theta(x_0-y_0) {G_0}^{-+}_{ij}(x,y) + \theta(y_0-x_0) {G_0}^{+-}_{ij}(x,y), \\
{G_0}^{--}_{ij}(x,y) &=  i \langle {\rm \bar{T}} \phi_i(x) \phi_j(y) \rangle^{(0)} =
\theta(x_0-y_0) 
{G_0}^{+-}_{ij}(x,y) + \theta(y_0-x_0){G_0}^{-+}_{ij}(x,y),
\end{align}
where the sub or superscript $(0)$ denotes the free field correlation
functions.

\subsection{One-loop logarithmic divergence}

As mentioned in the introduction, it has previously been shown that the logarithmic IR divergences to the power spectrum of inflationary perturbations can lead to large effects, which ultimately requires a non-perturbative treatment \cite{Sloth:2006az,Sloth:2006nu,seery1,seery2,Urakawa:2008rb}. Let us first briefIy review how these IR divergences manifest themselves in the present case. The perturbative expression for the two-point function is
\beq
\bra \phi_i(\eta_0,k)\phi_i(\eta_0,k')\ket = 
\int_{ctp}\mathcal{D}{\phi_i}^+\mathcal{D}{\phi_i}^- \phi^+_i(\eta_0,k)\phi^+_i(\eta_0,k') e^{\left[i\int_{t_{in}}^t dt' \int d^3x \sqrt{-g} (\mathcal{L}[{\phi_i}^+]-\mathcal{L}[{\phi_i}^-])\right]}~.
\eeq
We can then calculate the one-loop contribution by expanding the 
exponential to first order and perform the required Wick contractions 
\bea\label{1loop}
\bra  \phi_i(\eta_0,k)\phi_i(\eta_0,k') \ket &=& (2\pi)^3
\delta^{(3)}(\vec{k}+\vec{k}')
\int_{-\infty}^{\eta_0}\frac{d\eta}{\eta^4H^4} 2\Im  \left[G^{-+}_0(k;,\eta_0,\eta)G^{-+}_0(k;,\eta_0,\eta)\right] \nonumber\\ & & \times \lambda\, G^{-+}_0(x,x)\;.
\eea
On the right hand side we have suppressed the latin indices on the Green functions. The expression above has two logarithmic divergences. The most severe
 comes from
\beq
G^{-+}_0(x,x) \propto \int \frac{dk}{k} \mathcal{P}_{\phi}(k), 
\eeq
where the power spectrum of the scalar fields, $\mathcal{P}_{\phi}(k)$, is scale independent  on super-horizon scales in de Sitter space. The integral therefore diverges logarithmically with the lower limit of the $k$ integral. Naturally one may take the lower limit of the integral to be given by the largest scale which can fit inside the in inflationary patch at the beginning of inflation $k = a_i H_i$. This implies that the logarithmic divergency will scale like the physical size of the inflationary patch or equivalently like the total number of e-folding of inflation. 
The other logarithmic divergence comes from the time integral in Eq.~(\ref{1loop}). It depends on $k\eta_0$ and scales like the number of e-foldings left of inflation after the mode $k$ left the horizon. This time-dependence simply reflect that the perturbation variable is not conserved (the associated 
comoving curvature
perturbation in the case of only adiabatic modes will be conserved, see Refs.
\cite{seery1,seery2}).
Thus, the exact non-perturbative two-point function is expected to depend 
on $\eta_0$.
In order to deal with the former more severe IR divergency non-perturbatively, we will apply the Two-Particle-Irreducible formalism.

\subsection{Two-Particle-Irreducible formalism} 

From the generating functional defined above, we can define the generating functional of connected Green functions $W=-i\log Z$, and by a Legendre transformation, we can then compute the effective action $\Gamma$, which is the generating funtional of One-Particle-Irreducible (1PI) Green functions.

However, in the non-perturbative regime where quantum fluctuations are large, the 1PI effective action is insufficient because it only gives the dynamics of the mean field $\phi_c=\langle \phi \rangle$ and does not describe the evolution of the variance $\langle \phi^2\rangle$. In the 2PI approach the functional derivative of the effective action by the Green functions yields an equation for the variance, called the \textit{gap} equation, just like the dynamics of the classical field is given by the functional derivative of  the effective action by the classical mean field.

In the 2PI approach, one generalizes the generating function to include a non-local source term \cite{Cornwall:1974vz,Ramsey:1997qc}
\bea
& &Z[J_i^a, K_{ij}^{ab}] = \nonumber\\
&& \prod_{i,a} \int_{\text{{\tiny ctp}}} D \phi^i_a
\exp \Bigg[ i \biggl( 
S\left[ \phi_i^a \right] 
+ \int d^4 x \sqrt{-g} c_{ab} J_i^a \phi_j^b \delta^{ij}
\nonumber \\ &&
+ \frac{1}{2} \int d^4 x  \sqrt{-g} \int d^4 x'  \sqrt{-g'} c_{ab} 
c_{cd} K^{ac}_{ij}(x,x')
\phi_k^b(x) \phi_l^d(x') \delta^{ik} \delta^{jl} 
\biggr) \Biggr],
\eea
where summation over repeated indices is implicit, the matrix $c_{ab}$ is defined as $c^{++}=-c^{--}=1$, $c^{+-}=c^{-+}=0$, and 
\beq
\mathcal S[\phi_i^a] = \int d^4 x \sqrt{-g} (\mathcal{L}[{\phi_i}^+]-\mathcal{L}[{\phi_i}^-]).
\eeq
The effective action is a double Legendre transform of $W(J,K)$. One defines the classical field and the two-point function as 
\beq
{\phi_c}_i^a (x) = \frac{c^{ab}}{\sqrt{-g}}\frac{\delta W}{\delta J_j^b(x)}
\delta_{ij},
\eeq
\beq
{\phi_c}_i^a (x) {\phi_c}_j^b (x') + G_{ij}^{ab}(x,x') =
2\frac{c^{ac}}{\sqrt{-g}}\frac{c^{bd}}{\sqrt{-g'}} \frac{\delta W}{\delta 
K_{lm}^{cd}(x,x')} \delta_{ik} \delta_{jl}. 
\eeq
Now one can eliminate
$J_i^a$ and $K_{ij}^{ab}$ in terms of ${\phi_c}_i^a$ and $G_{ij}^{ab}$, 
and define the 2PI effective action as a double
Legendre transform of $W$,
\bea
& &\Gamma[\phi_c, G] = W[J, K] -
\int d^4 x \sqrt{-g} c_{ab} J_i^a {\phi_c}_j^b \delta^{ij} 
\nonumber \\ &&~~ -
\frac{1}{2} \int d^4 x \sqrt{-g} \int d^4 x' \sqrt{-g'} c_{ab} 
c_{cd} K_{ij}^{ac}(x,x') 
\Bigl[
 G_{kl}^{bd}(x,x') + {\phi_c}_k^b(x) {\phi_c}_l^d (x') \Bigr] 
\delta^{ik} \delta^{jl},
\eea
In a Feynman series expansion of $\Gamma$, one then obtains

\bea
\Gamma[\phi_c,G] &=& 
S[\phi_c] 
- \frac{i}{2} \ln  
\text{det} \left[ G_{ij}^{ab} \right] 
\nonumber \\ &&
+ \frac{i }{2} \int d^4 x \sqrt{-g} \int  d^4 x' 
\sqrt{-g'} {G_{0}^{-1}}^{ij}_{ab}(x',x)
G_{ij}^{ab}(x,x') + \Gamma_2[\phi_c,G],
\eea
where
\begin{equation}
i {G_{0}^{-1}}^{ij}_{ab}(x,x') = \frac{1}{\sqrt{-g}} 
\frac{\delta^2  S}{
\delta \phi_i^a (x) \phi_j^b (x')}  \frac{1}{\sqrt{-g'}}\;.
\end{equation}
After shifting the field around the background field $\phi\to \psi$ in the classical action $S[\phi]$ order to define a fluctuation field $\psi=\phi-\phi_c$, the new action becomes $S[\phi_c+\psi]$, and $\Gamma_2$ is then given by all closed 2PI graphs (all 2-point graphs which does not become open by opening two propagator lines), with all the propagator lines given by $G$ and vertices given by the shifted action.
 
 The mean field equation
 \beq
\left. \frac{\delta \Gamma[\phi_c,G]}{\delta {\phi_c}_i^a(x)}\right|_{
\phi_c^+ = \phi_c^- \equiv 
\phi_c} = 0
\eeq
and the gap equation
\beq
\left.\frac{\delta \Gamma[\phi_c,G]}{\delta G_{ij}^{ab}(x,y)}\right|_{ 
\phi_c^+ = \phi_c^- \equiv \phi_c} = 0,
\eeq
 can then be calculated to yield \cite{Aarts:2002dj}
 \beq
-\left(\square_x + m^2 + \frac{\lambda}{2N}
\left[ \phi_c^2(x) + G_{jj}^{++}(x,x) \right] \right) {\phi_c}_i(x)  = \frac{\lambda}{N} {\phi_c}_j(x)G_{ji}^{++}(x,x)  - \frac{\delta \Gamma_2[\phi,G]}{\delta {\phi_c}_i(x)},
\eeq
and 
\bea
-\left[\square_x + m^2 + \frac{\lambda}{2 N}{\phi_c}^2(x) \right]
G_{ij}^{++}(x,x') &=& 
\frac{\lambda}{ N} {\phi_c}_i(x){\phi_c}_k(x)G_{kj}^{++}(x,x')  \\ 
&+ & i \int d^4 z \Sigma_{ik}^{++}(x,z;\phi_c,G) G_{kj}^{++}(z,x') + i \delta_{ij} 
\frac{\delta(x-x')}{\sqrt{-g}} \nonumber
\eea
with 
\beq
\Sigma_{ij}^{++}(x,y;\phi_c,G) \equiv 
2i\frac{\delta \Gamma_2[\phi_c,G]}{\delta G_{ij}^{++}(x,y)}.
\eeq

\section{Solutions in the large $N$ expansion}

The leading order large $N$ approximation corresponds to a self-consistent mean field approximation, where the infinite hierarchy of Schwinger-Dyson equations for the $n$-point functions are truncated to a closed system of two dynamical equations of just the one-point function $\left<\phi_i(x)\right>$ and the two-point function $\left<\phi_i(x)\phi_j(y)\right>$. Since no irreducible higher order correlation functions appear in the large $N$ limit, it is equivalent to a Gaussian approximation for the density matrix.

In order to calculate the leading $N$ limit of the mean field and gap equations, we note that a trace over the $O(N)$ indices gives a factor of $N$, while the vertices each contribute a factor $1/N$. One can for simplicity rescale the field and composite field operators \cite{Cornwall:1974vz,Ramsey:1997qc}
\bea \label{resc} 
{\phi_c}_i^a(x) & \to & \sqrt{N} \phi_c^a(x), \\
G_{ij}^{ab}(x,x') & \to & G^{ab}(x,x')\delta_{ij}, \\
\psi_i^a(x) &\to & \psi^a(x)
\eea
in order to obtain the mean field equation to leading order in the large $N$ expansion
 \beq
-\left(\square_x + m^2 + \frac{\lambda}{2}
\left[ \phi_c^2(x) + G^{++}(x,x) \right] \right) {\phi_c}(x)  = 0.
\eeq
Using that to leading order in large $N$
\beq
\Gamma_2^{LO}[G]= -\frac{\lambda}{8N}\int d^4 x\sqrt{-g}\left[G_{ij}^{++}(x,x)G_{kl}^{++}(x,x)-G_{ij}^{--}(x,x)
G_{kl}^{--}(x,x)\right]\delta^{ij}\delta^{kl},
\eeq
one can also obtain the gap equation for the rescaled fluctuation Green function
\bea \label{gln}
-\left(\square_x + m^2 + \frac{\lambda}{2}
\left[ \phi_c^2(x) + G^{++}(x,x) \right]  \right)
G^{++}(x,x') =   i 
\frac{\delta(x-x')}{\sqrt{-g}}.
\eea
In the equation above, we have omitted the appropriate counter terms needed to cancel the UV divergences, since in the following we will only concern ourselves with the IR properties of the equation.

\subsection{Large wavelength approximation}

The first step in solving the gap equation for $G^{++}(x,x')$ is to solve for $G^{++}(x,x)$. 
Once we have an analytical expression for $G^{++}(x,x)$, we can insert it in the equation above and solve for $G^{++}(x,x')$.
We now assume that the $N$ scalar fields are evolving during a de Sitter stage characterized by
a constant Hubble rate $H$. 
In the large wavelength approximation, we can approximate the d'Alembertian by $3H\p_t$. Assuming further that initially $\phi_c=0$, and that it will stay small due to the $\phi\to -\phi$ symmetry of the Lagrangian, in the massless case the gap equation will at late times simplify to 
\bea
-3H\p_t G^{++}(x,x) = \lambda
\left[G^{++}(x,x)\right]^2+ i2 
\frac{\delta(0)}{\sqrt{-g}}~,
\eea
where the extra factor of two is a symmetry factor, which accounts for the difference in taking the derivative of $G(x,x')$ before or after  taking the coincidence limit. 
The appearance of  the ambiguous $\delta(0)$ implies that, in order to solve this equation, we must supplement the equation with a boundary condition. In the Hartree approximation in Ref.~\cite{Starobinsky:1994bd}, it was obtained by requiring that one recovers the correct linear growth of the variance in the limit $\lambda \to 0$ . 

One may however also observe, that the contribution from the IR modes is captured by introducing a smearing with a characteristic size corresponding to a fixed physical distance of the order of the horizon scale, and this gives rise to an additional term in the equation above governing the time evolution of the smeared two point function, as was also discussed in Ref.~\cite{Starobinsky:1994bd}. 
Below, we will instead solve for  $G^{++}(x,x)$ using the Fokker-Planck equation and then stick the result back into the original gap equation and then subsequently solve for $G^{++}(x,x')$.

\subsection{Fokker-Planck equation}

The stochastic approach embodies the idea that the IR part of the scalar field may be
considered as a classical space-dependent
stochastic field satisfying a local Langevin-like equation. 
The stochastic noise terms arise from the quantum fluctuations which become
classical at horizon crossing and then contribute to the background \cite{Starobinsky:1994bd}. 
The expectation value of any function $F[\phi]$ of the coarse-grained field is determined by 
\beq
\bra F[\phi] \ket = \int d\phi F(\phi)P(\phi,t).
\eeq
Using that in the massless case, the classical Lagrangian corresponding to the rescaled field in eq.~(\ref{resc}) is 
\beq
\mathcal{L}= \frac{N}{2} \left(\p_{\mu}\phi\p^{\mu}\phi -\frac{\lambda}{4}\phi^4\right)~,
\eeq
one can derive
a Fokker-Planck equation describing how the probability of scalar field values
at a given spatial point evolves with time
\beq
\frac{\p}{\p t}\left[P(\phi,t)\right] = \frac{\lambda}{6H}\frac{\p}{\p\phi}\left[\phi^3P(\phi,t)\right]+\frac{H^3}{8\pi^2}\frac{\p^2}{\p\phi^2}\left[P(\phi,t)\right].
\eeq
Now, let us consider the expectation value of the 2$n$th power of the field. Following Ref.~\cite{Tsamis:2005hd}, one can differentiate the expectation value of $\phi^{2n}$ and use the Fokker-Planck equation to obtain
\beq \label{fpv}
\frac{\p}{\p t}\left<\phi^{2n}\right> = \frac{n(2n-1)H^3}{4\pi^2}\left<\phi^{2n-2}\right>-\frac{n\lambda}{3H}\left<\phi^{2n+2}\right>  ,
\eeq
if one assumes that the field vanishes at the boundaries of the field integration. Like in Ref.~\cite{Tsamis:2005hd} one can define
\beq
\al \equiv \frac{1}{4\pi^2}\ln a,\qquad \bar\lambda\equiv 4\pi^2\lambda/3,
\eeq
in order to obtain the differential recursion relation
\beq
\frac{\p}{\p\al}\left< \left(\frac{\phi}{H}\right)^{2n}\right> = n(2n -1)\left< \left(\frac{\phi}{H}\right)^{2n-2}\right> -n\bar\lambda\left< \left(\frac{\phi}{H}\right)^{2n+2}\right>\eeq

It can be solved iteratively to yield the perturbative solution of Ref.~\cite{Tsamis:2005hd}
\beq
\left< \phi^{2n}\right>  = (2n-1)!!\left(\frac{H^2}{4\pi^2}\ln a\right)^n \left[1-\frac{n}{2}(n+1)\frac{\lambda}{12\pi^2}\ln^2 a+\dots \right] .
\eeq
However, in the case we are interested in we can solve the equation exact. Using that in the large $N$ limit $\bra \phi^4\ket = \bra \phi^2 \ket^2$ (only the trace squared part contributes in large $N$), we obtain for $n=1$
\beq
\frac{\p}{\p\al}\left< \left(\frac{\phi}{H}\right)^{2}\right> =1 -\bar\lambda\left< \left(\frac{\phi}{H}\right)^{2}\right>^2,
\eeq
which has the solution 
\beq \label{sol1}
\left< \left(\frac{\phi}{H}\right)^{2}\right> = \frac{\textrm{Tanh}(\frac{\sqrt{\bar\lambda}}{4\pi^2}\ln a)}{\sqrt{\bar\lambda}}.
\eeq
This agrees with the asymptotic solution for $\ln a\to \infty$ given in Ref.~\cite{Starobinsky:1994bd}, and exemplifies how the logarithmic IR divergences can be resummed to yield a finite result. The solution for $\bra\phi^2\ket=G^{++}(x,x)$ can now be inserted back into the gap equation, which can now be solved for $G^{++}(x,x')$.

\subsection{Analytical solution to the gap equation}

Let us assume that $G^{++}(x,x)$ is given by Eq.~(\ref{sol1}) and insert that into the gap equation Eq.~(\ref{gln}) and solve for $G(x,x')^{++}$ in the large wavelength limit, with $x\neq x'$. Then we obtain an equation of the form
\beq
\left(\p_t+\frac{\lambda}{2}\frac{\textrm{tanh}(at)}{b}\right)G(t)=0,
\eeq
where $a=\sqrt{\bar\lambda}H/4\pi^2$ and $b=3\sqrt{\bar\lambda}/H$. The solution is 
\beq \label{gap1}
G(t)= A\left(\textrm{cosh}(at)\right)^{-\lambda/2ab}.
\eeq
The solution scales like $\exp[-(\sqrt{3\lambda}/12\pi)Ht]$ at large times. 
If we are only interested in the asymptotic late time behavior, there is a more illuminating way to proceed. At late times the variance $\left<\phi^2\right> \to H^2/\sqrt{\bar\lambda}$, and in this case the gap equation in Fourier space yields
\beq
\left(\p_t^2 +3H\p_t+\frac{k^2}{a^2}+\frac{\sqrt{3\lambda}}{4\pi}H^2 \right)G^{++}(t,t_0;k)=0,
\eeq
where we neglected only the gradient. Now the interpretation is clear. The non-perturbative effect is a non-perturbative regulating mass given by $m_{\rm np}^2 = \frac{\sqrt{3\lambda}}{4\pi}H^2$. We can solve the equation by the usual method of going to conformal time and rewrite the equation on the form  
\beq \label{gap2}
\left(\p_\eta^2+k^2-\frac{1}{\eta^2}\left[\nu^2-\frac{1}{4} \right]\right)\tilde G^{++}(\eta,\eta_0;k)=0,
\eeq
where 
\beq
\nu^2=\left(\frac{9}{4}-\frac{m_{\rm np}^2}{H^2}\right).
\eeq
We defined the rescaled Green function as $G=\tilde G/a$.
As a function of $\eta$ the two independent solution scales like $\sqrt{-\eta} H_{\nu}^{(1)}(-k\eta)$ and $\sqrt{-\eta}H_{\nu}^{(2)}(-k\eta)$. Thus the asymptotic late time behavior scales as $(-k\eta)^{1/2-\nu}$ at late times. Thus, instead of going to a constant, $G^{++}(\eta,\eta_0;k)$ is decaying at late times as $(-k\eta)^{m_{\rm np}^2/H^2}$ in the limit $\eta\to 0$. This is the same late time behavior as given by the solution in Eq.~(\ref{gap1}). The exponential decay in physical time of the retarded Green function, implies that the system has a finite correlation time, which is proportional to $1/\sqrt{\lambda}H$.

\section{Non-perturbative enhancement of non-Gaussianities}

Let us consider the four point function of the $O(N)$ field in Fourier space. The interaction vertex for the rescaled fluctuation field in eq.~(\ref{resc}) is $(\lambda/8N) \psi^4$. At tree-level, the four-point function is given by
\bea
\bra \psi_{k_1} \dots \psi_{k_4}\ket &=&  -i \frac{3\lambda}{N} 
\delta^{(3)}(\vec{k}_1+\dots +\vec{k}_4) \\
&\times&\int^{\eta}\frac{d\eta'}{(H\eta')^4}\left[G_0^{++}(\eta,\eta';k_1)\dots G_0^{++}(\eta,\eta';k_4)-G_0^{--}(\eta,\eta';k_1)\dots G_0^{--}(\eta,\eta';k_4)\right],\nonumber
\eea 
where $G_0$ is the Green function of the free field. The Green function of a free massless field in de Sitter is given by 
\beq
G_0^{-+}(\eta,\eta';k)= i\frac{H^2}{2k^3}(k\eta-i)(k\eta'+i)e^{ik(\eta'-\eta)}~.
\eeq
Expanding the real and complex part of the free field Green function to leading order in the large wavelength limit, one then obtains a expression for the four-point function, which is logarithmically divergent at late times. At late times the tree-level four-point function is 
(we simplified the expression isolating the leading terms) \cite{Bernardeau:2003nx}
\beq
\label{ii}
\bra \psi_{k_1} \dots \psi_{k_4}\ket \simeq   \frac{3\lambda}{N} 
\delta^{(3)}(\vec{k}_1+\dots +\vec{k}_4) H^4 \frac{k_1^3+k_2^3+k_3^3 +k_4^3}{24 k_1^3k_2^3k_3^3 k_4^3} (\gamma+\log(-k\eta)).
\eeq
The divergent log term is proportional to the number of e-folding left before
 the end of inflation after the mode has crossed the horizon, and for observationally interesting modes, 
it will be of order one\footnote{At any rate, one has to take into account that, when computing the 
trispectrum of the gauge-invariant isocurvature perturbations, the time dependence may differ from the one in Eq. (\ref{ii}).}.
Like for the two-point correlation function, one expects cumulative effects at one-loop coming from the IR.
It is interesting to know how this conclusion is 
altered when we take into account all of the resummed loop effects by using the non-perturbative slotution for the Green function. 

The effect of the non-perturbative Green function is to regulate the log-divergent integral by the non-perturbative mass $m_{\rm np}^2$. With the non-perturbative mass, the Green function at late times take the form
\beq\label{GHan}
G(\eta,\eta';k)= i\frac{\pi}{4}(\eta\eta')^{3/2}H^2H_{\nu}^{(1)}(-k\eta)H_{\nu}^{(2)}(-k\eta').
\eeq
If we expand the Hankel functions around zero, we can straightforwardly calculate the four-point function at late times. However, a less tedious way to proceed is to make an appropriate field redefinition.


\subsection{Change of basis and classical approximation}
Since we are interested in investigating the infrared properties of the inflaton propagator on super Hubble scales, where the dynamics can be described classically, it is convenient to make a rotation in field space, in order to make the comparison with classical field theory simpler. The change of basis is defined by\footnote{For alternative diagrammatic approaches, see Ref.~\cite{Musso:2006pt,Byrnes:2007tm}.} \cite{Meulen:2007ah}
\beq
\left(\begin{array}{c}\psi^{(1)} \\ \psi^{(2)}\end{array}\right)= \bf{R}\left(\begin{array}{c}\psi^+ \\ \psi^- \end{array}\right)~,\qquad \bf{R}\equiv \left(\begin{array}{cc}1/2 & 1/2 \\ 1  & -1 \end{array}\right)
\eeq
In the new basis the Lagrangian density $\mathcal{L}[\psi^+]-\mathcal{L}[\psi^-]$, with the background field put to zero for simplicity, becomes
\beq
\mathcal{L}[\psi^{(1)},\psi^{(2)}] = \sqrt{-g}\left( -\p_{\mu}\psi^{(1)}\p^{\mu}\psi^{(2)}-m^2\psi^{(1)}\psi^{(2)}-\frac{\lambda}{8N}\left(4(\psi^{(1)})^3\psi^{(2)}+\psi^{(1)}(\psi^{(2)})^3\right)\right)~,
\eeq
and the propagator matrix becomes
\beq
\bf{G}_K(x,y) = \bf{R}\bf{G}(x,y)\bf{R}^T = \left(\begin{array}{cc}iF(x,y) & G^R(x,y) \\ G^A(x,y) & 0\end{array}\right)~,
\eeq
with
\bea
F(x,y) &=& -\frac{i}{2}(G^{-+}(x,y)+G^{+-}(x,y))~,\\
G^R(x,y) &=& G^{++}(x,y)-G^{+-}(x,y) = \theta(x_0-y_0)(G^{-+}(x,y)-G^{+-}(x,y))~,\\
G^A(x,y) &=& G^{++}(x,y)-G^{-+}(x,y) = \theta(y_0-x_0)(G^{+-}(x,y)-G^{-+}(x,y))~,\\
\eea
Like in Ref.~\cite{Meulen:2007ah}, we represent the $\psi^{(1)}$ field with a full line and the
$\psi^{(2)}$ field with a dashed line. The Feynman rules for the two
point functions and the vertices are then

\begin{align}
\picb{
\Line(2,15)(44,15)
\Text(4,24)[lt]{$\tau_1$}
\Text(43,24)[rt]{$\tau_2$}
} 
\hspace{.0cm} & = F(k,\tau_1,\tau_2), \\
\picb{
\Line(2,15)(22,15)
\DashLine(22,15)(44,15){3}
\Text(4,24)[lt]{$\tau_1$}
\Text(43,24)[rt]{$\tau_2$}
} 
\hspace{.0cm} & = -i G^R(k,\tau_1,\tau_2) = -i
G^A(k,\tau_2,\tau_1), 
\label{propsfeyn} \\
\picb{
\DashLine(4,15)(22,15){3}
\Text(8,17)[lb]{$\tau_1$}
\Line(22,15)(22,30)
\Text(25,24)[lb]{$\tau_2$}
\Line(22,15)(22,0)
\Text(35,24)[lt]{$\tau_3$}
\Line(22,15)(40,15)
\Text(25,7)[lt]{$\tau_4$}
} & = -i \frac{3\lambda}{N} \, a^4(\tau_1) \delta(\tau_1-\tau_2) \delta(\tau_1-\tau_3) \delta(\tau_1-\tau_4),
\label{cvertex} \\  
\picb{
\DashLine(4,15)(22,15){3}
\Text(8,17)[lb]{$\tau_1$}
\Line(22,15)(22,30)
\Text(25,24)[lb]{$\tau_2$}
\DashLine(22,15)(22,0){3}
\Text(35,17)[lb]{$\tau_3$}
\DashLine(22,15)(40,15){3}
\Text(25,7)[lt]{$\tau_4$}
} & = -i \frac{3\lambda}{4N} a^4(\tau_1) \delta(\tau_1-\tau_2) \delta(\tau_1- \tau_3)\delta(\tau_1- \tau_4).
\label{qvertex} \end{align}
When a two point function is attached to a vertex, the corresponding
time has to be integrated over. A closed loop corresponds with an
integral over spatial momentum $\int d^3p/(2 \pi)^3$. 

The tree level contribution to the
equal time two point function  
\begin{equation}
\int d^3 x \; e^{-i \mathbf{k} \cdot \mathbf{x}}
\langle\psi(\tau,\mathbf{x}) \psi(\tau,\mathbf{0}) \rangle ,
\label{def2point} 
\end{equation}
is given by
\begin{equation}
\picb{
\Line(2,15)(44,15)
\Text(4,24)[lt]{$\tau$}
\Text(43,24)[rt]{$\tau$}
} 
\hspace{1.5cm} F (k,\tau,\tau).
\end{equation}
There is no contribution with the $G^R$ two point function because
that vanishes for equal times. 

It has been shown in Ref.~\cite{Meulen:2007ah}, that on large scales, in the classical approximation, we can ignore the vertices in Eq.~(\ref{qvertex}) with three dashed lines.

\subsection{Non-perturbative tri-spectrum}

In the classical limit where we can ignore the vertex with three dashed lines, the tri-spectrum will now simply be given by
\bea
\bra \psi_{k_1} \dots \psi_{k_4}\ket &=&  -i \frac{3\lambda}{N} \delta(k_1+\dots +k_4) \\
&\times&\int^{\eta}\frac{d\eta'}{(H\eta')^4}F(\eta,\eta';k_1)F(\eta,\eta';k_2)F(\eta,\eta';k_3)G^R(\eta,\eta';k_4)+perm,\nonumber
\eea 
where $perm$ symbolize all permutations of $1,2,3,4$.

Using
\bea
F (x,y)&=& \Im\left(G^{-+}(x,y)\right)\\
G^R(x,y) &=& \theta(x_0-y_0)2\Re\left(G^{-+}(x,y)\right),
\eea
one can find the leading order contributions to the Green functions in the small momenta limit, by expanding the Hankel funtions in Eq.~(\ref{GHan}) in the small argument limit. In this way one obtains  the infrared limit of the propagators
\beq
F(k;\tau_1,\tau_2) \approx \frac{H^2}{2k^3}(-k\tau_1)^{\ep}(-k\tau_2)^{\ep},
\eeq
and 
\beq
G^R(k;\tau_1,\tau_2) \approx \theta(\tau_1-\tau_2)\frac{H^2}{k^3}\left[(-k\tau_1)^{\ep}(-k\tau_2)^{3-\ep}-(-k\tau_2)^{\ep}(-k\tau_1)^{3-\ep})\right]~.
\eeq
For convenience we have defined $\ep\equiv m_{\rm np}^2/3H^2$.
With these expressions, the non-perturbative solution for the tri-spetrum in the large wavelength limit becomes
\bea
\bra \psi_{k_1} \dots \psi_{k_4}\ket &=&  \frac{3\lambda}{N} \delta^{(3)}(\vec{k}_1+\dots +\vec{k}_4) H^4 \frac{k_1^3+k_2^3+k_3^3 +k_4^3}{24 k_1^3k_2^3k_3^3 k_4^3}\frac{2\ep-3}{8\ep^2-6\ep}\sum_{i<j<l}(k_i\eta)^{2\ep}(k_j\eta)^{2\ep}(k_l\eta)^{2\ep}\nonumber\\
&\approx& 
\frac{1}{2\ep} \frac{3\lambda}{N} \delta^{(3)}(\vec{k}_1+\dots +\vec{k}_4) H^4 \frac{k_1^3+k_2^3+k_3^3 +k_4^3}{24 k_1^3k_2^3k_3^3 k_4^3},
\eea
where in the last step we took the limit of small $\ep$.  One may worry if we should take the limit $\eta\to 0$ before taking the limit of small $\ep$. However, in the integral above the non-vanishing contribution comes from the super-horizon part, which implies that in the expression above $k\eta=-(k/H)\exp(-Ht)$ is exponentially suppressed by the number of e-foldings left of inflation after the mode $k$ has left the horizon. For physically interesting modes the number of e-foldings left of inflation after horizon crossing is less than about $60$. Thus, taking immediately $\eta \to 0$ is not entirely physical, and the limit of small $\ep$ in the equation above is a good approximation when $\ep\lesssim 1/60$.

Inserting the value $\ep=m_{\rm np}^2/3H^2=\sqrt{3\lambda}/{12\pi}$ finally yields 
\beq
\bra \psi_{k_1} \dots \psi_{k_4}\ket \approx  
\frac{ 36\pi\sqrt{3\lambda}}{N}\delta^{(3)}(\vec{k}_1+\dots +\vec{k}_4) H^4 \frac{k_1^3+k_2^3+k_3^3 +k_4^3}{24 k_1^3k_2^3k_3^3 k_4^3},
\eeq
so effectively the tri-spectrum experience a non-perturbative enhancement by a factor $12\pi/\sqrt{\lambda}$ when $\lambda$ is small. Notice also that
the higher-loop corrections to the trispectrum are suppressed in the IR by the
fact that the propagator $F(k,\tau_1,\tau_2)$ scales like $k^{2\delta-3}$
in the IR.

\section{Non-perturbative corrections to the curvature perturbations}

Now, let us as a final example, discuss an example similar to the one discussed by Weinberg in Ref.~\cite{Weinberg:2005vy}, where the interaction between the curvature perturbation $\zeta$ and a set of test scalar fields $\phiß_n$ was considered. If we assume that the test scalar fields are only gravitationally coupled to the inflaton, the part of the Lagrangian containing the test scalar fields and their gravitational interactions is, similarly to in Ref.~\cite{Weinberg:2005vy}, given by (with $8\pi G=1$)
\bea
& &
{\mathcal L}=\frac{a^3}{2}e^{3\zeta}\Bigg[-2{\mathcal N}V(\phi_n)+{\mathcal N}^{-1}\sum_n\Big(\dot{\phi}_n-{\mathcal N}^i\partial_i\phi_n\Big)^2-{\mathcal N}a^{-2}e^{-2\zeta}[\exp{(-\gamma)}]^{ij}\sum_n\partial_i\phi_n\partial_j\phi_n\Bigg]\;,\nonumber\\ & &{ }
\eea
where the spatial part of the mertic in this gauge is given by
\begin{equation}
g_{ij}=a^2e^{2\zeta}[\exp{\gamma}]_{ij} ,~~~~~\gamma_{ii}=0,~~~~~\partial_i\gamma_{ij}=0\;.
\end{equation}
where $\gamma_{ij}({\bf x},t)$ is a gravitational wave amplitude, and $\zeta({\bf x},t)$ is the curvature perturbation. The other components of the metric are given by the lapse and the shift functions, $\mathcal{N}$, and $\mathcal{N}^i$, in the usual ADM approach
\beq
g_{00}=-{\mathcal N}^2+g_{ij}{\mathcal N}^i{\mathcal N}^j\;,~~~~~~~g_{i0}=g_{ij}{\mathcal N}^j\;.
\eeq
Now we can expand ${\mathcal N}$ in a sequence in perturbation theory ${\mathcal N}=1 + \al^{(1)}+\al^{(2)}+\dots$. Taking the potential of the $O(N)$ invariant test scalar $V(\phi)= (\lambda/8N)(\phi\cdot\phi)^2$, then from expanding the exponential, we would naively conclude that there should be an interaction term of the type 
\beq
{\mathcal L}_{\zeta^2\phi^4} = \frac{9}{16N}\lambda a^3\zeta^2(\phi\cdot\phi)^2\;,
\eeq
although one can not trust the slow-roll order of that term. Since the Lagrangian for the curvature perturbation should vanish in pure de Sitter limit, the Lagrangian should be suppressed by some power of the slow-roll parameter \cite{Maldacena:2002vr,Jarnhus:2007ia}. To find the right slow-roll order of the interaction term, it is convenient to use instead the variable $\zeta_n$ \cite{Maldacena:2002vr,Jarnhus:2007ia}, which is defined as $\epsilon \zeta_n=Q$, where $Q$ is the inflaton fluctuation in the uniform curvature gauge and $\epsilon$ is the first slow-roll parameter. Writing the Lagrangian in the uniform curvature gauge to second order in $\zeta_n$, one finds a slow-roll suppressed contribution of the form\footnote{The term proportional to $\alpha^{(2)}_{\zeta_n}V(\phi)$ cancels out after applying the background equations. The leading surviving contribution is proportional to $(\alpha^{(1)}_{\zeta_n})^2V(\phi)$, where $\al^{(1)}_{\zeta_n}=(1/4)\epsilon^2\zeta_n^2$ was given in Ref.~\cite{Maldacena:2002vr} (while $\al_{\zeta_n}^{(2)}$ was computed in Ref.~\cite{Sloth:2006az,Seery:2006vu}).}
\beq
{\mathcal L}_{\zeta^2_n\phi^4} =\frac{1}{32N}\lambda a^3\epsilon^2\zeta_n^2(\phi\cdot\phi)^2\;.
\eeq
From the interaction above, we can calculate the contribution to the two-point function of curvature perturbations
\bea
\bra {\zeta_n}\zeta_n \ket &=& (2\pi)^3\int_{-\infty}^{\eta_0}\frac{d\eta}{\eta^4H^4}  2\left[G^>_{\zeta_n}(k;,\eta_0,\eta)G^>_{\zeta_n}(k;,\eta_0,\eta)
-G^<_{\zeta_n}(k;,\eta_0,\eta)G^<_{\zeta_n}(k;,\eta_0,\eta)\right]\nonumber\\
& & \times \frac{1}{8}\epsilon^2\lambda NG^>_{\phi}(x,x)G^>_{\phi}(x,x)\;.
\eea
Using 
\beq
G^>_{\zeta_n}(k;,\eta_0,\eta) = i \frac{1}{\epsilon}U_k(\eta_0)U_k^*(\eta)\;,
\eeq
and 
\beq
U_k(\eta) = \frac{iH}{k\sqrt{2k}}(1+ik\eta)e^{-ik\eta}\;,
\eeq
we obtain
\bea
\bra {\zeta_n}\zeta_n \ket &=&(2\pi)^3\frac{\lambda N}{2} \int_{-\infty}^{\eta_0}\frac{d\eta}{\eta^4H^4}\Im\left[U_k^2(\eta_0)U_k^{*2}(\eta)\right]G^>_{\phi}(x,x)G^>_{\phi}(x,x)\nonumber\\
&=& (2\pi)^3 \frac{\lambda N}{12k^3}\left[\frac{20}{3}+\frac{2}{3}\textrm{Ci}(-2k\eta_0)+\dots\right]G^>_{\phi}(x,x)G^>_{\phi}(x,x)\;.
\eea
If we insert the asymptotic value $G^>_{\phi}(x,x)=G^{++}(x,x)\approx H^2/2\pi\sqrt{\lambda}$, which we calculated in section 4, then we obtain (remember we are taking $8\pi G=1$)
\bea
\bra {\zeta_n}\zeta_n \ket &=&\frac{4\pi N H^4}{6 k^3}\left[\frac{20}{3}+\frac{2}{3}\textrm{Ci}(-2k\eta_0)+\dots\right]\;.\nonumber\\
&=&  \frac{8 \pi^2}{3k^3}\epsilon N  H^2 {\mathcal P}_{\zeta}\left[\frac{20}{3}+\frac{2}{3}\textrm{Ci}(-2k\eta_0)+\dots\right]\;,
\eea
where ${\mathcal P}_{\zeta}$ is the tree-level power spectrum of curvature perturbations and the logarithmic running is because $\zeta
_n$ is not conserved. However, we can evaluate the expression at horizon crossing when Ci$(-k\eta_0)\sim 1$ and make a straightforward transformation into the conserved $\zeta$. Because we are considering initially a two-loop correction, we see that there is an suppression of an additional $H^2$ which typically will make the contribution small. Nevertheless this simple analytic solution illustrates the behavior of non-perturbative effects, and, like in the previous section, we find a parametric enhancement in the coupling constant $\lambda$.  One may then speculate about the non-perturbative behavior of other more realistic models with one-loop contributions, which will scale with one power of $H^2$ less and where we know that the perturbative one-loop result can already be significant. However, a detailed study of such more complicated cases are beyond the scope of the present paper and is left for future work.

\section{Summary and conclusions}
In this paper we have analyzed the properties of the cosmological perturbations during a de Sitter stage of a set of
scalar fields subject to an $O(N)$ symmetry with quartic
interaction and in the limit of large $N$. In such a regime the theory can be solved exactly 
keeping control of the IR cumulative effects  which appear in perturbation theory and are
proportional to the total number of e-folds of the inflationary stage.  
One finds that the
correlation functions are IR finite. 
In particular, we have computed the resulting trispectrum, which is the interesting quantity signalling the presence
of NG's. We have   shown to parametrically differ from the tree-level value. 
Its amplitude is  suppressed only by the square root of the quartic coupling, 
thus manifesting 
the presence of non-pertrubative effects. 
In 
our study we have limited ourselves to the computation of the scalar field correlation functions; 
we have not addressed the issue of relating these correlations to those of the truly gauge-invariant
(iso-)curvature perturbations.  However, this step is automatic through the 
$\delta N$ formalism \cite{sk1,sk2,Sasaki:1995,Lyth:2004gb,Lyth:2005fi} once the time-evolution of the universe is known. Finally, we have computed the
non-perturbative 
corrections to the two-point correlator of the comoving perturbation
coupled to a seto of $N$ fields, i.e. the toy model
proposed by Weinberg \cite{wein2} to show the presence of one-loop IR
divergences. We have shown that the non-perturbative corrections are
again parametrically different from the one-loop result. 
Our findings may be considered as a first step towards the understanding of the
IR cumulative effects which plague at the loop-level cosmological scalar perturbations when the theory
is self-interacting and a superlarge box is assumed.

\overskrift{Acknowledgments}

\noindent  The authors are grateful to N. Bartolo, M. Pietroni, 
S. Matarrese  for useful conversations and comments. We would also like to thank the anonymous referee for some useful comments on the first version. This research was supported in part by the European Community's Research
Training Networks under contracts MRTN-CT-2004-503369 and MRTN-CT-2006-035505.

\end{document}